\documentclass[11pt,a4paper, twoside]{book}
\pdfoutput=1
\usepackage[T1]{fontenc}
\usepackage[adobe-utopia]{mathdesign}  		
\usepackage{a4wide}                     	
\usepackage[english]{babel}             	
\usepackage{amsmath}                    	
\usepackage{amsfonts}						
\usepackage{makeidx}                    	
\usepackage{url}                        	
\usepackage{graphicx}                   	
\usepackage[small,bf,hang]{caption}     	
\usepackage{xspace}                     	
\usepackage{subfig}
\usepackage[latin1]{inputenc}           	
\usepackage{float}                      	
\usepackage{flafter}                    	
\usepackage[nottoc]{tocbibind}		    	
\usepackage{fancyhdr}                   	
\usepackage[Gray,squaren,thinqspace,thinspace]{SIunits} 
\usepackage{hyperref}
\usepackage{appendix}
\author{T.J. Huisman}
\title{Single photon experiments}
\usepackage{perpage}									
		
\makeindex                             
\usepackage{longtable}         			
\usepackage{pdfpages}

\begin{document}

\begin{titlepage}

\fontsize{11pt}{16pt}\selectfont

\begin{center}

\vspace{4cm}

\textit{\textbf{\begin{Large}Characterization of a Quantum Light Source Based on Spontaneous Parametric Down-Conversion\end{Large}}}\\
\vspace{0.5cm}

\vspace{1.5cm}

\textbf{Thomas J. Huisman,${^*}$ Simon R. Huisman,${^*}$ Allard P. Mosk, Pepijn W.H. Pinkse${^\dagger}$}\\
\begin{small}
\textit{MESA+ Institute for
Nanotechnology, University of Twente, PO Box 217, 7500 AE Enschede, The Netherlands}
\\
\textit{$^*$ Both authors contributed equally}\\
\textit{$^\dagger$Corresponding author: p.w.h.pinkse@utwente.nl, www.adaptivequantumoptics.org}\\
Enschede, February 2013
\end{small}
\end{center}

\vspace{2.5cm}
We have built a quantum light source capable of producing different types of quantum states. The quantum light source is based on entangled state preparation in the process of spontaneous parametric down-conversion. The single-photon detection rate of 8$\cdot 10^5$ s$^{-1}$ demonstrates that we have created a bright state-of-the-art quantum light source. As a part of the characterization we measured two-photon quantum interference in a Hong-Ou-Mandel interferometer.\\
\end{titlepage}

\thispagestyle{empty}
\cleardoublepage

\tableofcontents 

\chapter{Quantum light source at the few-photon level} \label{photon_source_chapter}

All quantum optical experiments need a light source for producing the desired states. We have chosen a quantum light source based on the  non-linear optical process of spontaneous parametric down-conversion (SPDC). SPDC is a process that was for the first time experimentally demonstrated in the seventies of last century\cite{DCBurnham_parametric_photon_pairs}. In SPDC a photon with a high energy is down converted into an entangled photon pair. Nowadays SPDC is a commonly used technique for creating heralded single photons\cite{SACastelletto_heralded_photon_source}, entangled photon states\cite{PGKwiat_new_engtangled_source}, squeezed states\cite{NJain_bridge_single_squeezed} and other quantum states\cite{AZavatta_single_cohterent_state,EBimbard_state_engineering}. Given the capability of producing many diverse quantum states in a well-defined spatial mode at a high rate, makes SPDC an ideal process for diverse experiments.\\

In this chapter we discuss the requirements and results of the quantum light source that we have built. We start by introducing the working principles of the non-linear optical processes used in our setup. Sequentially, we present the quantum light source that we have built. We continue with characterizing the performance of the quantum light source by measuring the detection rates of single- and two-photon Fock states. We end this chapter with the conclusions. 

\section{Non-linear optical processes} \label{photon_prep_section}

In our setup we make use of three types of non-linear optical processes: second-harmonic generation (SHG), optical parametric amplification (OPA) and spontaneous parametric down-conversion (SPDC). We make use of non-linear crystals that have a large second-order susceptibilities ($\chi ^{(2)}$), meaning that they have a strong non-linear response to external electric fields. Second-order susceptibilities allow three-wave mixing, meaning that two waves of different frequencies can be mixed to make a third wave of another frequency, or the other way around. We continue with discussing SHG, as it makes a good starting point for explaining the other two non-linear processes.\\

\begin{figure}[tb]
\centering
\subfloat[]{\includegraphics[scale=1.2]{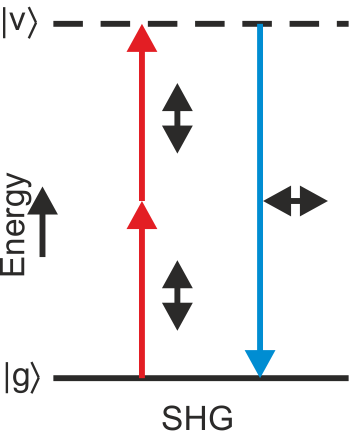}\label{theory_SHG_fig}} 
\hspace{2cm}
\subfloat[]{\includegraphics[scale=1.2]{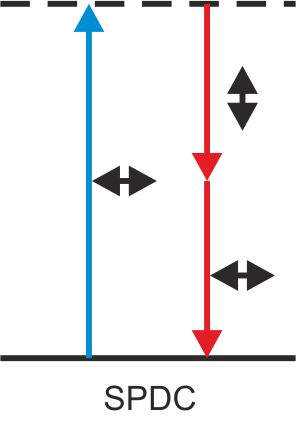}\label{theory_SPDC_fig}} 
\caption{\textbf{Energy diagrams of two non-linear optical processes.} The vertical colored arrows represent the frequency of the light as used in our experiment. The symbols next to these arrows indicate the polarization of the light as used in our experiment, the polarization is either vertical or horizontal with respect to the optical axis of the crystal. $\mid g \rangle$ and $\mid v \rangle$ indicate the ground and virtual energy level involved in the non-linear process. \textbf{(a)} Second-harmonic generation, two identical photons are annihilated and a frequency-doubled photon is created. \textbf{(b)} Spontaneous parametric down-conversion, a photon with a high frequency is annihilated and two photons with lower frequencies and orthogonal polarizations are created.}
\label{theory_non_linear_optics_fig}
\end{figure}

In our experimental setup we use SHG to convert our laser wavelength into a suitable wavelength for the quantum light source. SHG is a process in which the frequency of light is doubled; two photons of a long wavelength are annihilated while a photon with half the wavelength of the original photons is created. An energy diagram of SHG is presented in figure \ref{theory_SHG_fig}. 
 All non-linear optical processes fulfill the phase-matching conditions, in case for SHG these are given by\cite{fundamentals_of_photonics}:
\begin{eqnarray}
\hbar \omega _1 + \hbar \omega _2  = \hbar \omega _3 \label{energy_phase_match}\\
\hbar \bf{k} _1 + \hbar \bf{k} _2  = \hbar \bf{k} _3 \label{momentum_phase_match}
\end{eqnarray}
where $\omega$ is the frequency of light, $\bf{k}$ is the wave vector and the indices indicate the photons involved in the process. Equation (\ref{energy_phase_match}) represents energy conservation and equation (\ref{momentum_phase_match}) represent momentum conservation. For SHG energy conservation essentially prescribes the wavelength of the created photon (3) as function of the other two photons (1,2). Momentum conservation is achieved by using non-linear crystals with convenient dispersion and birefringent properties. For SHG we use type-I phase matching in a BBO crystal, meaning that light used to drive this process is orthogonally linearly polarized compared to the generated frequency-doubled light.\\
 
We use OPA for alignment purposes. OPA is a process in which light of a short wavelength is used to amplify light of a long wavelength. SPDC on the other hand is essential in our setup since we use it to create entangled photons. In SPDC a pump photon is spontaneously down converted into two photons called a signal and an idler photon, see also figure \ref{theory_SPDC_fig}. The created photons are correlated in position and anticorrelated in energy\cite{WPGrice_spectral_info_SPDC} and entangled in polarization\cite{PGKwiat_new_engtangled_source}. We use a type-II periodically poled KTP non-linear crystal for SPDC, meaning that the created photons will appear in orthogonal linear polarizations. In this chapter we use the entanglement of the photon pairs as a way to herald the presence of a single photon; by detecting one of the created photons you are sure another photon with orthogonal polarization was created. Creating entangled photons is a rare process and is essential for the experiment shown in the consecutive chapter.

\subsection{Why use non-linear optics to create single photons} \label{why_SPDC_section}

As we mentioned in the introduction of this chapter, we use spontaneous parametric down-conversion (SPDC) to create photon pairs, because it can produce many diverse quantum states in a well-defined spatial mode at a high rate. Alternatives for approximating a single photon source are for example\cite{GSBuller_single_photon_generation}: emmission of single hydrogen-like atoms\cite{MKeller_single_photons_ion_trap}, color centers in diamond\cite{CKurtsiefer_diamond_photons} and quantum dots\cite{PMichler_single_quantum_dot,SStrauf_many_single_photons}. From these examples, ref. \cite{SStrauf_many_single_photons} has the highest single-photon detection rate, $4\cdot 10^6$ s$^{-1}$. Our setup has a single-photon detection rate of about $5\cdot 10^5$ s$^{-1}$, whereas detection rates lower than 1$\cdot 10^5$ s$^{-1}$ are common for the other single-photon sources. All of these processes have their advantages and disadvantages. A drawback of using SPDC is that the photons are not created on demand. This problem can be resolved partly by using a pulsed light source, allowing the production of photons only during the pump pulse duration\cite{WPGrice_spectral_info_SPDC,SACastelletto_heralded_photon_source}. Besides single-photon pairs, SPDC can also produce higher order number pairs. In a simplified notation the state produced by SPDC can be written as\cite{HHansen_PhD_thesis}:
\begin{equation}
\mid \psi _{\rm{SPDC}} \rangle =C_{\rm{norm}} \left( \mid 0_s 0_i \rangle + \gamma \mid 1_s 1_i \rangle + \frac{1}{2} \gamma ^2 \mid 2_s 2_i \rangle + ... + \frac{1}{n!} \gamma ^n \mid n_s n_i \rangle \right) \label{SPDC_approx_eq}
\end{equation}
were $C_{\rm{norm}}$ is a normalization constant and $s$ and $i$ stand for signal and idler respectively. $\gamma$ is a prefactor depending on the parameters of the crystal and the pump beam\cite{NJain_MSc_thesis}, which can be given as $\gamma \propto A_{\omega _1} \chi ^{(2)}$, where $A_{\omega _1}$ is the amplitude of light that is being spontaneously down converted. $\gamma$ is chosen $\ll$1 in order to avoid producing more than one photon in a mode. As ref. \cite{HHansen_PhD_thesis} points out, triggering on $n$ photons does not exclude the probability of detecting $m>n$ photons, however this is not likely compared to detecting $n$ photons. In this way besides a heralded single-photon source, also a heralded $n$-photon source can be approximated with SPDC. SPDC is a very reliable source of single photons and it is unique in the diversity of quantum states it can produce.\\

\section{Experimental setup}

\begin{figure}[tb]
\centering
\includegraphics[scale=1.2]{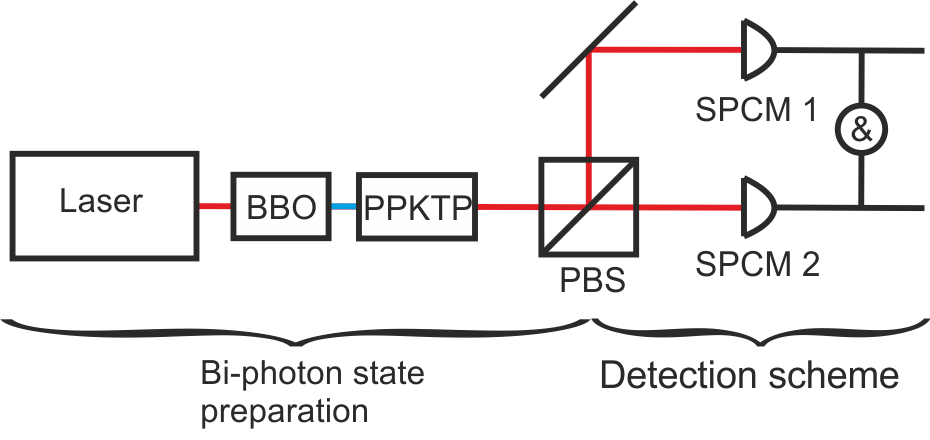}
\caption{\textbf{Overview of the quantum light source.} Laser light is frequency doubled with a BBO non-linear crystal (BBO). The frequency-doubled light is used to pump a periodically poled KTP non-linear crystal (PPKTP), producing photon pairs with orthogonal polarization. The photons are separated with a polarizing beam splitter (PBS) and detected with single-photon counting modules (SPCM 1 and SPCM 2).} \label{simplified_setup}
\end{figure}

Figure \ref{simplified_setup} provides an illustration of the experimental setup.
Our setup is based on an existing setup\cite{NJain_MSc_thesis,Simon_internship_report}. A mode-locked titanium sapphire laser emits pulses with a center wavelength at 790 nm, a repetition rate of 82 MHz and a pulse width around 0.4 ps. Most of the light is frequency doubled in a single pass through a 5-mm-long BBO crystal (BBO) cut for 790 nm. Remaining laser light is removed using two spectral filters (Semrock FF01-440/SP-25). The frequency-doubled light is then coupled into a 2-mm-long periodically poled KTP (PPKTP) crystal causing spontaneous parametric down-conversion (SPDC), after which the frequency-doubled light is removed using a spectral filter (Semrock BLP01-635R-25). The photon pairs created with SPDC pass through this filter and are separated using a polarizing beam splitter (PBS). The two photons can be guided to separate single-photon counting modules (SPCM 1 and SPCM 2: PerkinElmer SPCM-AQRH-14 and PerkinElmer SPCM-AQRH-13 respectively) using single-mode fibers which make it possible to collect the single photons from a well-defined spatial mode. A detailed description of the single-photon source can be found in appendix \ref{single_photon_setup_appendix}.\\

\subsection{Laser requirements} \label{light_source_section}

For efficient creation and detection of photon pairs, the light source needs to fulfill a good transverse mode quality for good mode matching. Thereby a moderately high-power level should be used to drive the process of spontaneous parametric down-conversion\cite{NJain_MSc_thesis,Simon_internship_report,HHansen_PhD_thesis}. Also the wavelength of the light should be such that either directly or by conversion with non-linear optics the light can be used to efficiently create and detect photon pairs. Wavelengths around 700 nm can give more than 70\% detection efficiency with a silicon-based single-photon counting module\cite{APD_sheet} and seem therefore ideal.\\

A pulsed light source is preferred over a continues-wave light source, as a way to approximate the preference for photons on demand. In practice a pulsed light source will be required in order to have a sufficient intensity to drive all the non-linear optical processes. A pulsed light source has a certain spectral width inherently related to the temporal pulse width. In order to perform several quantum interference experiments, e.g. the one described in the next chapter, a narrow-band light source should be used. From theory\cite{SACastelletto_heralded_photon_source} (see also section \ref{HOM_interference}) and other experimental setups\cite{NJain_MSc_thesis,Simon_internship_report,HHansen_PhD_thesis} follows that pulse widths of a few picoseconds are optimal. Thereby the repetition rate should not exceed the detection rate of the system\cite{NJain_MSc_thesis,Simon_internship_report,HHansen_PhD_thesis}.\\ 

The laser that best approaches all of these condition and which is available for this project at our university, is a Spectra Physics Tsunami mode-locked Ti:Sapphire laser pumped with a Millennia Xs laser. The output beam of the Ti:Sapphire laser is slightly elliptical (with a ratio of 1.6 for the horizontal over vertical beam width) and has typically an average output power of approximately 750 mW. The center wavelength of the laser light is 790 nm and the full-width-at-half-maximum of the spectrum is typically around 3 nm. The corresponding temporal pulse width as measured with an autocorrelator is around 0.4 ps. This means that our pulses are slightly chirped because the time bandwidth product, that is the spectral pulse width multiplied with the temporal pulse, is around 0.6 in our case while ideally this is 0.32 for our laser\cite{OSvelto_principles_lasers,Tsunami_manual}. The laser is specified to have a repetition rate of 82 MHz, which does exceed the detection rate of our system by about a factor of two. More details about the working principle of the laser can be found in appendix \ref{laser_appendix}.

\section{Characterization of the quantum light source}

\subsection{Measured heralded single-photon detection rates} \label{single_photon_results_section}

A full characterization of the count rates for the heralded single-photon source is summarized in table \ref{early_results_photon_source_table}.
\begin{table}[tb] 
\begin{center}
  \begin{tabular}{ | p{4cm} | p{2.3cm} | p{2.3cm} |  p{2.3cm} |}
    \hline
    & SPCM 1\newline counts \newline $\left[ 10^3 \rm{\ s}^{-1} \right]$ & SPCM 2\newline counts \newline $\left[ 10^3 \rm{\ s}^{-1} \right]$ & Coincidence counts  \newline $\left[ 10^3 \rm{\ s}^{-1} \right]$ \\ \hline  
    Dark counts & 0.12 $\pm$ 0.05 & 1.2 $\pm$ 0.2 & 0.00  $\pm$ 0.00\\
    Only entrance of \newline SPCM's blocked & 5 $\pm$ 1 & 11 $\pm$ 1 & 0.05 \\
    Everything operational & 747 $\pm$ 3 & 502 $\pm$ 5 & 140 $\pm$ 1\\    
    \hline    
    \end{tabular} 
  \caption{\textbf{Full characterization of count rates of the heralded single-photon source.}  The $\pm$ symbol indicates an estimate for the statistical error margin in which the counts were observed.}
  \label{early_results_photon_source_table}
  \end{center}
\end{table}
We observed photon-detection rates of 5.02$\cdot 10^5$ s$^{-1}$ and 7.47$\cdot 10^5$ s$^{-1}$ in the two separate modes. Thereby we observed a coincidence rate of 1.40$\cdot 10^5$ s$^{-1}$. Ideally the counts of the SPCM's and the coincidence counts are equal. However, the coincidence counts are lower than the SPCM counts, due to the limited efficiencies of the SPCM's and losses. The difference in count rates on the separate SPCM's are due to different detection efficiencies. To our knowledge, we can consider the count rates presented in table \ref{early_results_photon_source_table} being state-of-the-art. For example a very comparable setup detects around 1.00$\cdot 10^5$ single photons each second\cite{NJain_MSc_thesis,SRHuisman_single_photon_tomography}, while many other setups based on SPDC have lower count rates due to the use of non-linear crystals with a weaker non-linearity. The data obtained in table \ref{early_results_photon_source_table} were measured with a pump power of 77 mW.\\

For consecutive measurements we improved the setup, mainly the beam quality of the frequency-doubled light was improved. However, we did not do a full characterization of the count rates again because we modified the setup for other types of measurements. What we can observe in the count rates from the improvements nowadays is summarized in table \ref{recent_results_photon_source_table}.
\begin{table}[tb] 
\begin{center}
  \begin{tabular}{ | p{4cm} | p{2.3cm} | p{2.3cm} |  p{2.3cm} |}
    \hline
    & SPCM 1\newline counts \newline $\left[ 10^3 \rm{\ s}^{-1} \right]$ & SPCM 2\newline counts \newline $\left[ 10^3 \rm{\ s}^{-1} \right]$ & Coincidence counts\newline $\left[ 10^3 \rm{\ s}^{-1} \right]$ \\ \hline  
    Dark counts & 0.096 $\pm$ 0.010 & 0.765 $\pm$ 0.020 & 0.000 $\pm$ 0.000 \\
    Only entrance of \newline SPCM's blocked &  0.110 $\pm$ 0.014 &  0.805 $\pm$ 0.024 & 0.000 $\pm$ 0.000\\
   Everything operational & & 800 $\pm$ 2 & \\    
    \hline    
    \end{tabular} 
  \caption{\textbf{Characterization of count rates of the heralded single-photon source after  improving the setup.} The $\pm$ symbol indicates the standard deviation of 10 measurements.}
  \label{recent_results_photon_source_table}
  \end{center}
\end{table}
Clearly in our improved setup we are able to suppress the number of false counts almost to the level of dark counts on the detectors. Also we see an increase of counts on SPCM 2, while the intensity of the pump light used is lowered to 66 mW for the measurements in table \ref{recent_results_photon_source_table}. 

\subsection{Estimation of the two-photon Fock-state detection rate}

\begin{figure}[tb!]
\centering
\includegraphics[scale=1.2]{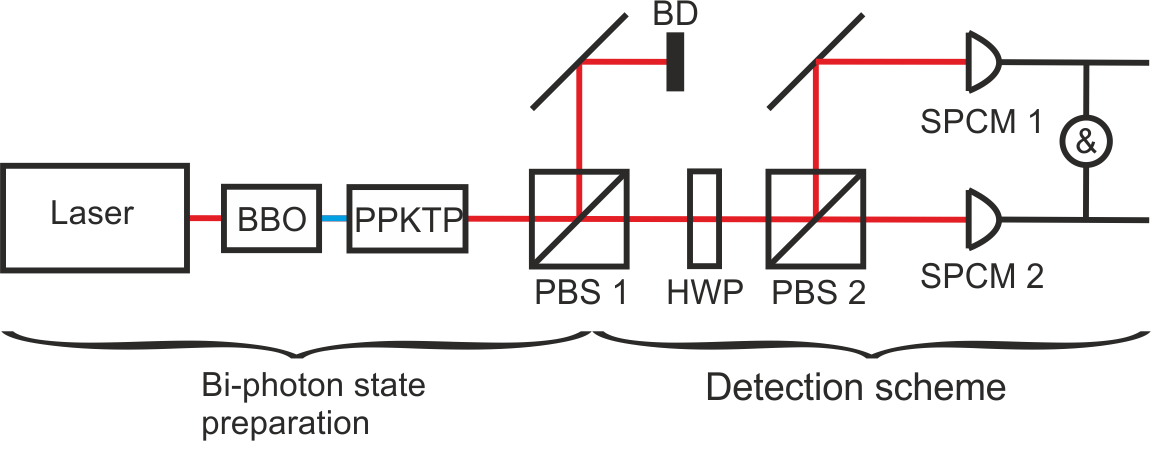}
\caption{\textbf{Overview of the setup used for estimating the two-photon Fock-state detection rate.} A laser, BBO non-linear crystal (BBO) and a periodically poled KTP non-linear crystal (PPKTP) create entangled photon pairs. The photons of these pairs are separated with a polarizing beam splitter (PBS 1). The reflected photons are directed to a beam dump (BD). The transmitted photons are guided to a half-wave plate (HWP) and a polarizing beam splitter (PBS 2). The outputs of PBS 2 are directed to single-photon counting modules (SPCM 1 and SPCM 2).} \label{simplified_setup_part_2}
\end{figure}

\begin{figure}[tb!]
\centering
\subfloat[]{\includegraphics[scale=0.6]{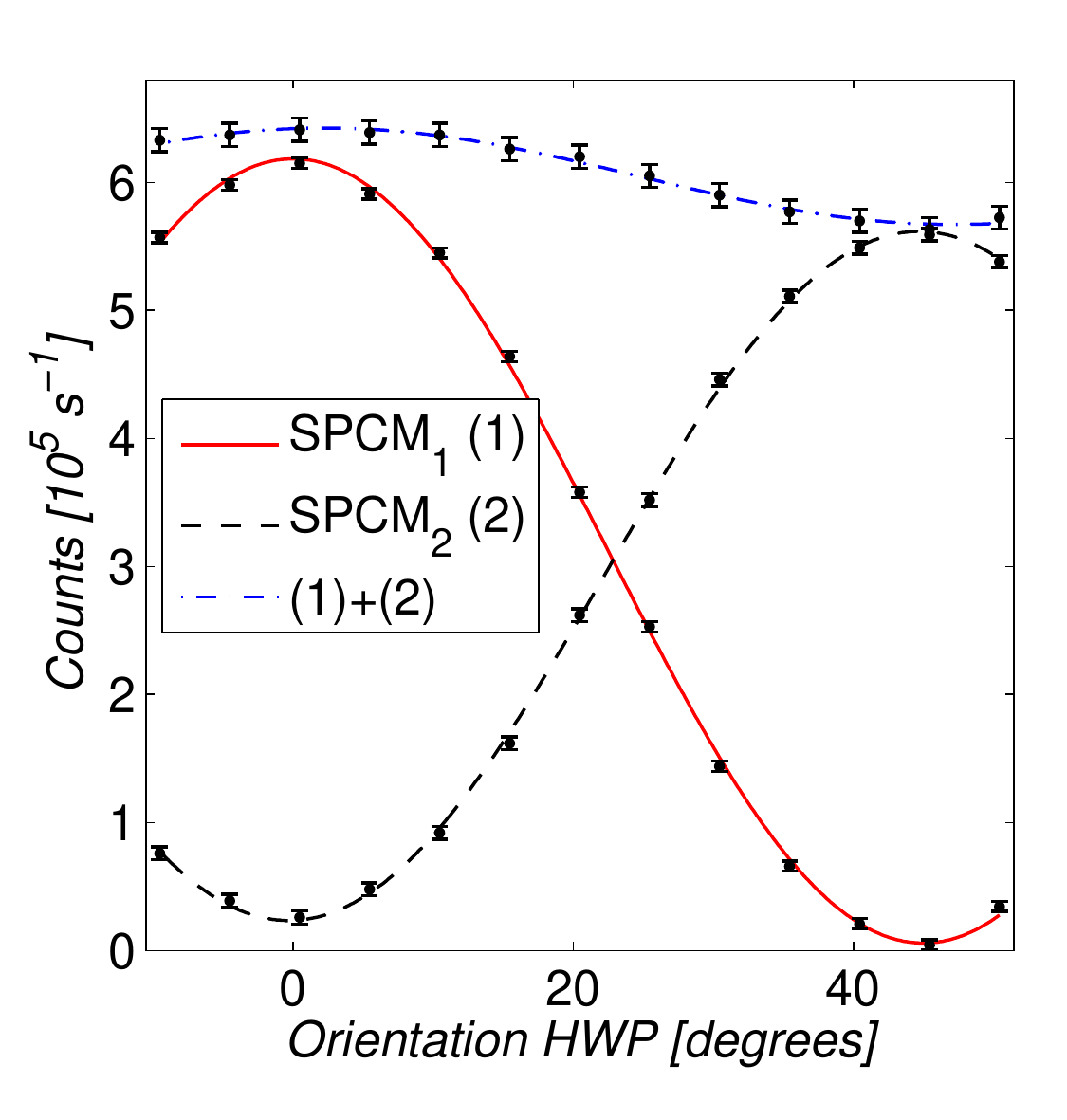}\label{2photon_apd_fig}} 
\hspace{0.5cm}
\subfloat[]{\includegraphics[scale=0.6]{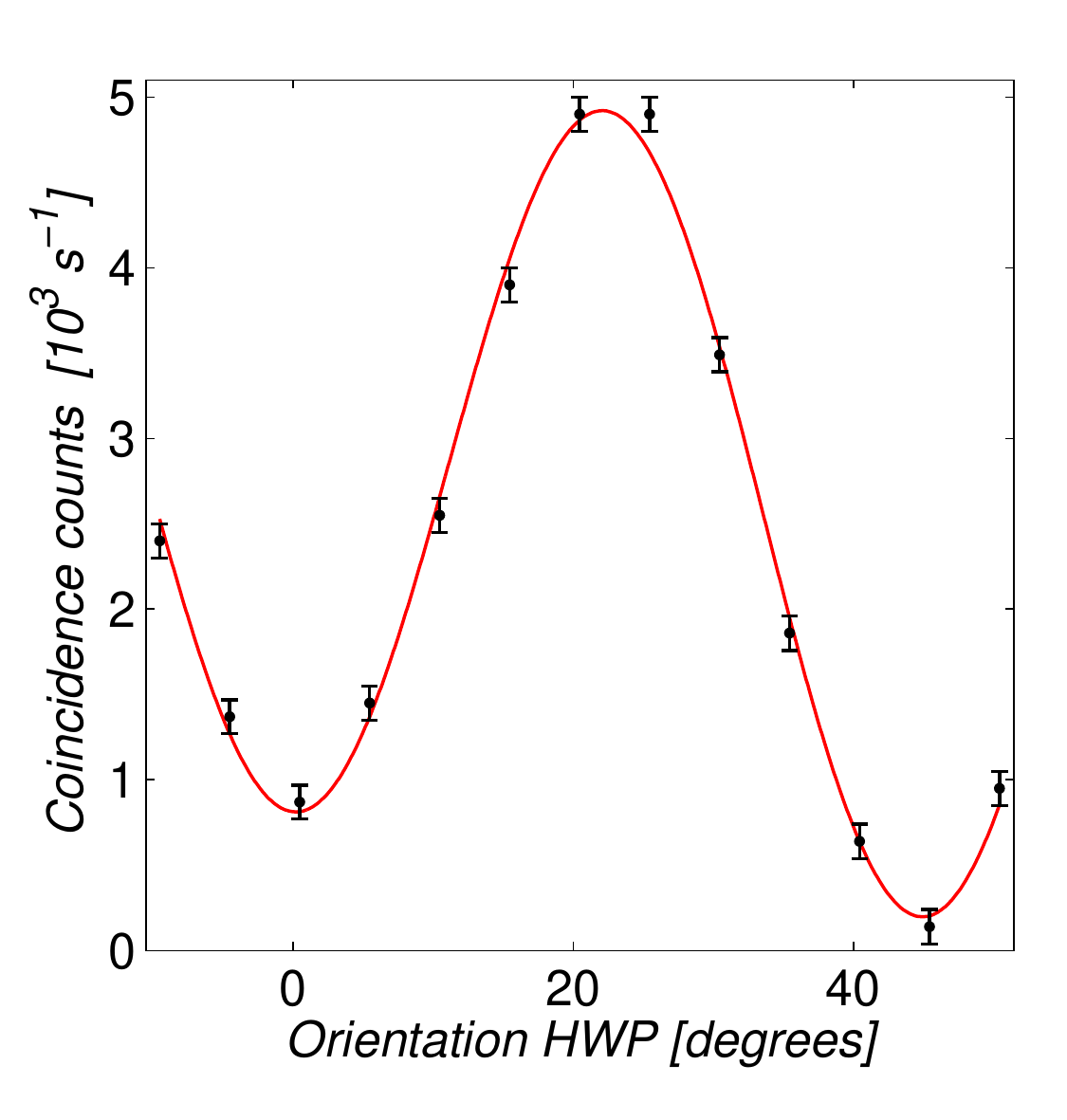}\label{2photon_coin_fig}} 
\caption{\textbf{Obtained results for estimating  the two-photon Fock-state detection rate} \textbf{(a)} The count rates of the individual SPCM's are depicted together with the total number of counts observed, all as function of the orientation of the half-wave plate (HWP) in figure \ref{simplified_setup_part_2}. The error margins are the maximum fluctuations in the observed counts. \textbf{(b)} The coincidence counts of the SPCM's are shown as function of the orientation of the half-wave plate in figure \ref{simplified_setup_part_2}. The error margins are the maximum fluctuations in counts observed.}
\label{2photon_figs}
\end{figure}

We also performed an experiment to estimate what the probability is to detect two-photon Fock-state pairs instead of a single-photon Fock-state pair (see equation (\ref{SPDC_approx_eq})). For this we blocked the reflected photons from PBS 1 in figure \ref{simplified_setup_part_2}. 
The photons transmitted through PBS 1 pass through a half-wave plate (HWP) and a polarizing beam splitter (PBS 2). After passing this polarizing beam splitter, the photons are focused onto two SPCM's. The half-wave plate in combination with a polarizing beam splitter act as a beam splitter with variable splitting ratio. From this small adjustment in the setup, we obtained the two graphs as shown in figure \ref{2photon_figs}.
In figure \ref{2photon_apd_fig} the counts of the separate SPCM's are depicted and the total number of counts is presented, all as function of the orientation of the HWP. The error bars in figure \ref{2photon_apd_fig} indicate the maximum fluctuations in counts observed and the lines are fits of the type
\begin{equation}
f_1+f_2\sin \left( \frac{2 \pi}{90} \theta + f_3 \right) \label{first_fitting_eq}
\end{equation}
were $f_{1-3}$ are fitting parameters and $\theta$ is the orientation of the half-wave plate in degrees. Equation (\ref{first_fitting_eq}) describes the intensity of light along one polarization axis as the half-wave plate is turned. Turning the half-wave plate 90 degrees does not alter the polarization axis, hence the period of 90 degrees in equation \ref{first_fitting_eq}. The offset in equation (\ref{first_fitting_eq}) ($f_1$) is included as a way to deal with non-ideal properties of the polarizing optics. With non-ideal properties of polarizing optics we mean for example the possibility of observing reflected light from a polarizing beam splitter, while ideally all the light should be transmitted meaning that the light was not in a perfectly defined linear polarization or the polarizing beam splitter is not capable of 100\% transmitting the light. As a result of these non-ideal properties of the polarizing optics, we can clearly observe in figure \ref{2photon_apd_fig} that the total number of counts depends on the orientation of the half-wave plate. Also, the total number of counts may vary due to detector efficiencies being different, which is indicated by different maximum count rates for the two SPCM's. The used pump power in figure \ref{2photon_figs} was 77 mW.\\ 

In figure \ref{2photon_coin_fig}, the coincidence counts as function of the orientation of the half-wave plate is given. The error bars indicate the maximum fluctuations in counts observed and the line is a fit of the type:
\begin{equation}
f_1+f_2\sin \left( \frac{2 \pi}{45} \theta + f_3 \right) - f_4\sin \left( \frac{2 \pi}{90} \theta + f_5 \right) \label{coin_fit_eq}
\end{equation}
The non-ideal properties of the polarizing optics and the different efficiencies of the detectors have the same orientation period of the half-wave plate as the polarization has, hence explaining the sine term with a frequency of $\frac{2 \pi}{90}$ in equation (\ref{coin_fit_eq}). The coincidence counts should be maximum when the polarization before PBS 2 in figure \ref{simplified_setup_part_2} is diagonal, which happens every half period of the half-wave plate, resulting in the sine term with a frequency of $\frac{2 \pi}{45}$ in equation (\ref{coin_fit_eq}). A clear peak can be observed in figure \ref{2photon_coin_fig} at 22.5 degrees, as expected, indicating the presence of the two-photon Fock-state. A rough estimate of the number of two-photon Fock-state would be to take the peak value of figure \ref{2photon_coin_fig} times two, because half of the two-photon states will be transmitted or reflected as pair at PBS 2 and will subsequently be counted once by one SPCM. This would amount to $9.8 \cdot 10^3$ two-photon Fock-state pairs each second. The total number of counts detected as presented in figure \ref{2photon_apd_fig} is about 6.05$\cdot 10^5$. This means that roughly 1.6\% of all observed photon counts is actually a two-photon Fock-state. Probably a more accurate estimations for the amount two-photon Fock states, would be to take the value of $f_2$ in equation (\ref{coin_fit_eq}) and multiply this value by four. In this way, effects of the background and other non-ideal properties are partly removed. This would amount to $9.0 \cdot 10^3$ two-photon Fock-state pairs each second, which would be 1.5\% of all observed clicks.  The two-photon Fock-state detection rate can be considered low compared to the detected rate of single-photon Fock states, but high enough to be considered for usage in future experiments.\\

There are several options to reduce the detection rate of two-photon Fock states. To suppress the two-photon Fock-state production rate, the pump power could be lowered so that this rate decreases quadratically with power. The production rate of single-photon Fock-state pairs on the other hand, decreases linearly with the pump power. From this it follows that the ratio of the production rates of the two photons over single photons decreases with decreasing pump power. To suppress the two-photon Fock-state detection rate in the heralded single-photon source, a 50:50 beam splitter could be placed at one of the output ports of PBS in figure \ref{simplified_setup} and direct both output ports of this 50:50 beam splitter to separate SPCM's. In this way it is possible to trigger partly on the two-photon Fock-state and remove these counts from the measurements. 

\section{Conclusions of the single-photon source}

In this chapter we described our  quantum light source. Detected single photon count rates exceeding 5$\cdot 10^5$ s$^{-1}$ and coincidence rates more than 100$\cdot 10^3$ s$^{-1}$ are common with our source. Based on a similar quantum light source\cite{NJain_MSc_thesis,SRHuisman_single_photon_tomography}, the observed count rates can be considered to be outstanding. We estimate that these count rates will be more than sufficient for most experiments that we have in mind. We also determined that roughly 1.5\% of the total observed counts are two-photon Fock states. This number can be considered low compared to the detected amount of single-photon Fock states, but high enough to be considered for usage in future experiments.

\chapter{Observing two-photon quantum interference} \label{HOM_chapter}

A decade after the experimental demonstration of entangled photon pair generation from spontaneous parametric down-conversion, Hong, Ou and Mandel demonstrated a two-photon interference effect\cite{CKHong_first_HOM} nowadays known as Hong-Ou-Mandel interference\cite{quantum_optics_Fox}. They showed that two photons incident on a 50:50 beam splitter show quantum interference, meaning that these photons interfere in a way that no semiclassical theory can explain\cite{JCGarrison_quantum_optics}. This interference is of great interest for time-resolved measurements on subpicosecond time scales\cite{AMSteinberg_photon_glass_HOM,AMSteinberg_photon_tunneling_HOM,DJPapoular_optical_tunneling_HOM} without the requirement of subwavelength stability\cite{CKHong_first_HOM,BZhao_robust_creation}.\\

In this chapter we investigate the feasibility of using our quantum light source for Hong-Ou-Mandel interferometry. In the first section of this chapter we explain Hong-Ou-Mandel interference and make quantitative predictions of how this interference influences the count rates in our setup. We continue with presenting the adjustments we made to our heralded single-photon source in order to measure Hong-Ou-Mandel interference. We end this chapter by showing our results and providing a summary.

\section{Hong-Ou-Mandel interference} \label{HOM_interference}

\begin{figure}[tb!]
\centering
\includegraphics[scale=1.2]{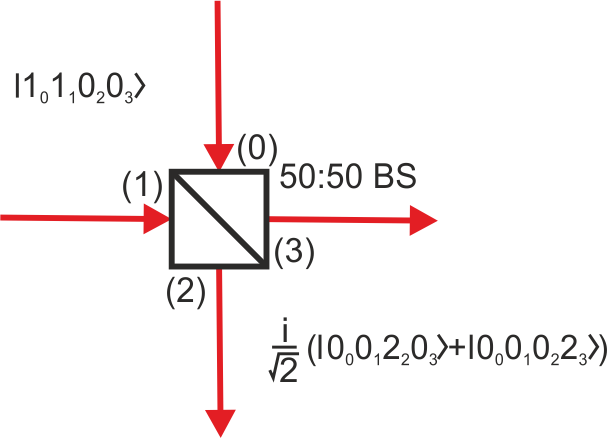}
\caption{\textbf{Two-photon quantum interference at a beam splitter.} Two identical photons incident on a 50:50 beam splitter from different ports will leave the beam splitter in pairs.} \label{HOM_principle}
\end{figure}

Hong-Ou-Mandel interference is based on two-photon quantum interference were two indistinguishable photons are incident on different ports ((0) and (1) in figure \ref{HOM_principle}) of an ideal 50:50 beam splitter. They can leave the beam splitter at the two remaining ports ((2) and (3) in figure \ref{HOM_principle}). 
With indistinguishable is meant that there is no way to tell by measurement, through which of the two input ports of the beam splitter the observed photons were transported. This means that the spectrum, polarization, and arrival time at the beam splitter for both photons should be identical. In this case both photons will exit the beam splitter from the same port\cite{quantum_optics_Gerry_Knight,JCGarrison_quantum_optics,quantum_optics_Fox}, as illustrated in figure \ref{HOM_principle}. For the beam splitter in figure \ref{HOM_principle} we can write two equations\cite{quantum_optics_Gerry_Knight}:
\begin{eqnarray}
\hat{a}_0 ^\dag=\frac{1}{\sqrt{2}} \left( \hat{a}_2 ^\dag+i\hat{a}_3 ^\dag\right) \nonumber\\
\hat{a}_1 ^\dag=\frac{1}{\sqrt{2}} \left( \hat{a}_3 ^\dag+i\hat{a}_2 ^\dag\right) \label{beam_splitter_interference}
\end{eqnarray} 
where $\hat{a}_i^\dag$ is the creation operator for mode $i$, indicating the different ports of the beam splitter in figure \ref{HOM_principle}. The input state in our case is given by $\mid \psi _{\rm{in}} \rangle = \mid 1_01_10_20_3 \rangle = \hat{a}_0 ^\dag \hat{a}_1 ^\dag \mid 0_00_10_20_3 \rangle$. From this, we can derive
\begin{eqnarray}
\mid \psi _{\rm{out}} \rangle &=& \frac{1}{\sqrt{2}} \left( \hat{a}_2 ^\dag  +i\hat{a}_3 ^\dag \right) \frac{1}{\sqrt{2}} \left( \hat{a}_3 ^\dag + i\hat{a}_2 ^\dag \right)  \mid 0_00_10_20_3 \rangle \nonumber\\
&=& \frac{i}{2}\left( \hat{a}_2 ^\dag \hat{a}_2 ^\dag + \hat{a}_3 ^\dag \hat{a}_3 ^\dag \right) \mid 0_00_10_20_3 \rangle \nonumber\\
&=& \frac{i}{\sqrt{2}} \left( \mid 0_00_12_20_3 \rangle + \mid 0_00_10_22_3 \rangle \right) \label{HOM_output_eq}
\end{eqnarray}
Equation (\ref{HOM_output_eq}) shows that the photons will leave one of the output ports of the 50:50 beam splitter in pairs. The probability for the photons to leave from separate ports of the beam splitter interferes destructively. The resulting quantum state is sometimes also referred to as a $N 0 0 N$-state with $N=2$, referring to the $N$-photon quantization, while detecting $0$-photons in the other output port. $N 0 0 N$-states allow to go beyond the diffraction limit and to do very sensitive interferometry, therefore these states are of high interest for quantum lithography\cite{ANBoto_quantum_lithography,YKawabe_quantum_diffraction_limit} and quantum metrology\cite{TNagata_beating_quantum_limit,IAfek_high_NOON}. The result of equation (\ref{HOM_output_eq}) does not hold if the photons are distinguishable. In the original Hong-Ou-Mandel interferometer and in our experiment, the arrival times of the two created photons before the 50:50 beam splitter were varied by introducing a path-length difference between these photons. Hong-Ou-Mandel interference can be observed by measuring both output ports of the beam splitter and detect coincidences, which should not occur for perfect Hong-Ou-Mandel interference at zero path-length difference.\\

\begin{figure}[tb!]
\centering
\subfloat[]{\includegraphics[scale=0.6]{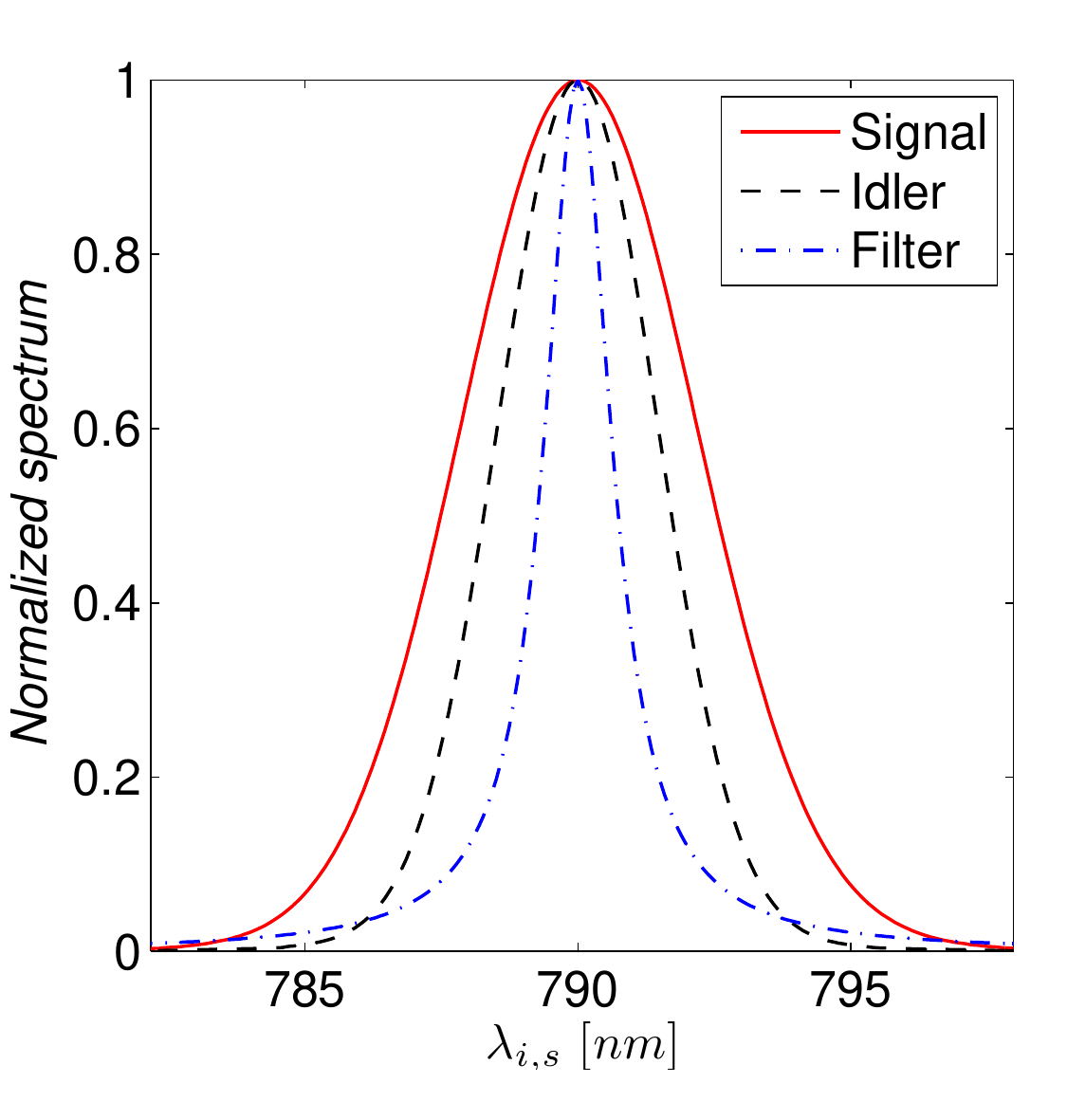}\label{theory_HOM_fig_1}} 
\subfloat[]{\includegraphics[scale=0.6]{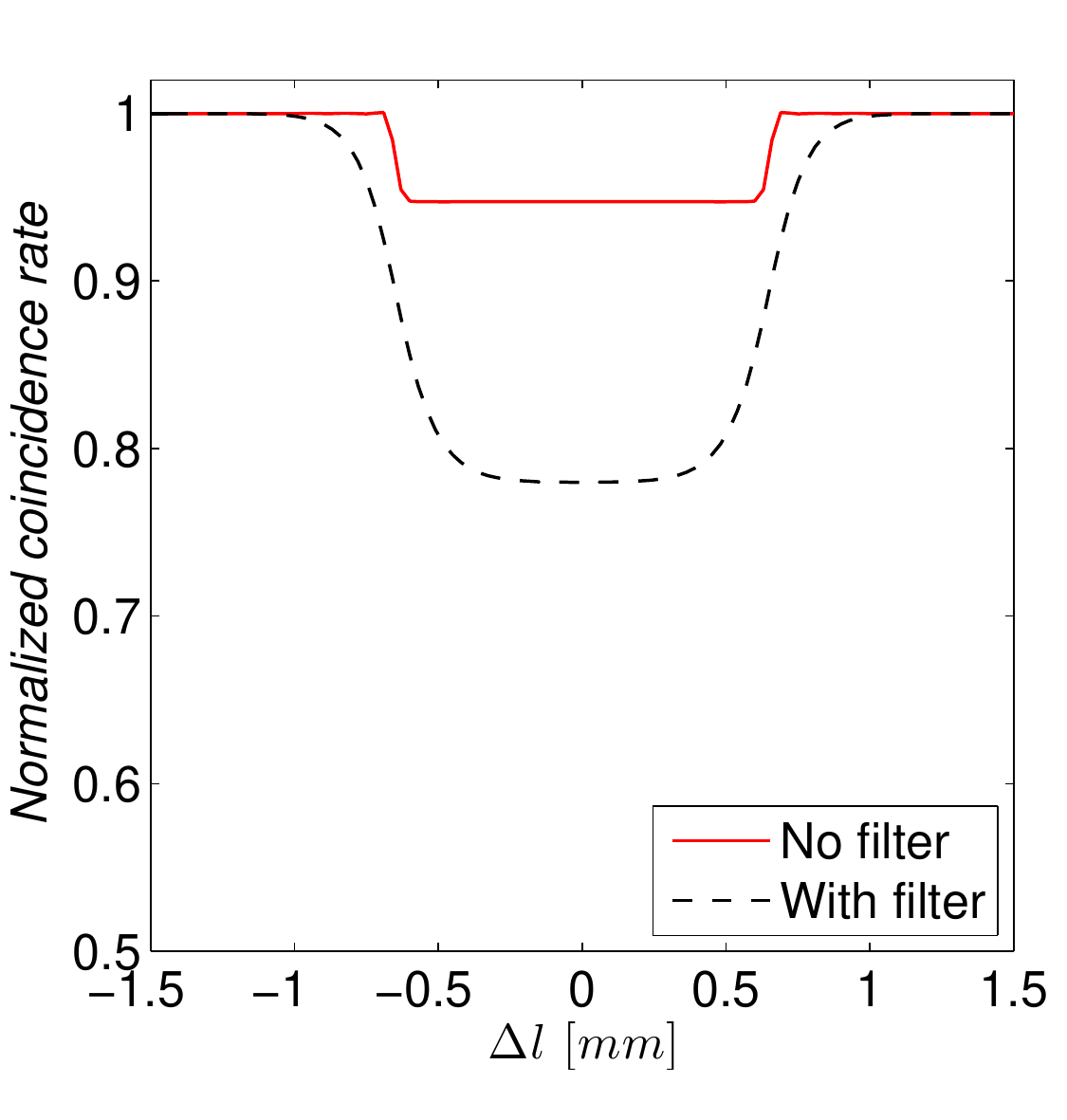}\label{theory_HOM_fig_2}} 
\caption{\textbf{Theoretical calculations for the Hong-Ou-Mandel experiment.} \textbf{(a)} Normalized photon spectra as function of the wavelength of the signal and idler photon. Also a normalized function of our spectral filter is presented. \textbf{(b)} Calculated coincidence rate in our experiment, with and without spectral filter.} \label{HOM_theory}
\end{figure}

Hong-Ou-Mandel interference depends on how distinguishable the two photons used are, and thus also on the spectral bandwidth in which these photons are produced. Two spectra are important in order to determine the spectrum of the created photons, the first being the spectrum of the pump laser, assumed to be a Gaussian and represented by $\alpha (\omega _s + \omega _t)$. The second spectrum results from the phase matching conditions for our non-linear crystal that creates our photon pairs, represented by $\Phi (\omega _s, \omega _i)$. The spectrum following from the phase matching conditions is described by the phase-matching function which is given by\cite{NJain_MSc_thesis,WPGrice_spectral_info_SPDC}:
\begin{equation}
\Phi (\omega _s, \omega _i) = \frac{\sin \left(\left[k_s(\omega _s)+k_i(\omega _i)-k_p(\omega _s+\omega _i)+2 \pi / \Lambda \right]L\right)}{\left[k_s(\omega _s)+k_i(\omega _i)-k_p(\omega _s+\omega _i)+2 \pi / \Lambda \right]L} \label{phase_match_eq}
\end{equation}
where the symbols are the same as used before in section \ref{photon_prep_section} and $\Lambda$ is the poling period of the crystal. Multiplying the phase-matching function with the spectrum of the pump laser gives the emitted two-photon spectrum $\left( \mid \alpha (\omega _s + \omega _t) \Phi (\omega _s, \omega _i) \mid^2 \right)$ of spontaneous parametric down-conversion (SPDC) as function of the wavelength of both photons. The spectrum of a photon is obtained by integrating the two-photon spectrum over the wavelength of the conjugate photon, resulting in figure \ref{theory_HOM_fig_1}. 
We observe in figure \ref{theory_HOM_fig_1} that the spectral width in which the photons are created is not the same for both photons, making both photons partly distinguishable. We can also observe from this figure that the spectral filter as used in our experiment (see appendix \ref{spectral_filter_section}) has a narrower spectral width than the bandwidth in which the photons are created. We can use this filter to make the spectrum of both photons more identical, either by applying the filter to both photon spectra like we did or by only filtering one photon spectrum. We approximated the properties of our non-linear crystal with the temperature-depended dispersion relations for KTP crystals taken from reference \cite{GGhosh_dispersion_KTP}. This dispersion relations is however not perfect for our non-linear crystal since the calculated phase-matching function (equation \ref{phase_match_eq}) would favor a center wavelength of 788.5 nm, while our crystal favors a center wavelength of 790 nm. To compensate this difference we multiplied the refractive index of the crystal in our model with a constant factor, this should approximate the properties of our non-linear crystal. We chose our other variables for calculating the spectra of the photons to be identical to the parameters used in the setup.\\

With the spectra of the photons we can calculate the result of a Hong-Ou-Mandel interferometer. For this we use equation (19) of reference \cite{WPGrice_spectral_info_SPDC}:
\begin{equation}
R_C (\Delta l) \propto \int \int d \omega _s \omega _t \mid \alpha (\omega _s + \omega _t) \mid ^2 \left[ \mid \Phi (\omega _s, \omega _t) \mid ^2 - \Phi (\omega _s, \omega _t) \Phi ^\ast (\omega _t, \omega _s) e^{-i (\omega _t - \omega _s) \Delta l / c} \right] \label{HOM_Rc_eq}
\end{equation}
where $c$ is the speed of light and $\Delta l$ is the relative delay between the input arms of the 50:50 beam splitter. $R_C$ is the coincidence rate, which is the rate at which photons are detected at the output ports of the 50:50 beam splitter. The first term in equation (\ref{HOM_Rc_eq}) is proportional to the total probability of observing coincidence counts. The second term oscillates rapidly with frequency ($\omega _t - \omega _s$) if $\Delta l$ is large, in which case this term has a negligible contribution to the coincidence rate. As $\Delta l$ approaches zero, this term contributes and the coincidence rate will be lowered. From equation (\ref{HOM_Rc_eq}) it is clear that the coincidence rate will decrease more when the spectrum of the signal and idler photons become identical. The calculated results of the Hong-Ou-Mandel interferometer for our setup is presented in figure \ref{theory_HOM_fig_2}. A clear dip of the Hong-Ou-Mandel interference can be observed, of which the width depends on the spectrum of the photons. The visibility of the dip is defined as $V=1-R_C(0)/R_C(\infty)$\cite{MTanida_good_HOM_SPDC}. From figure \ref{theory_HOM_fig_2} it is clear that filtering increases the visibility of the Hong-Ou-Mandel interference, because the spectra of the photons become identical. The normalized coincidence rates as given in figure \ref{theory_HOM_fig_2} do not depend on the number of photons detected and are therefore robust against time-independent losses and absorption. This is different when the measurement is suffering from detector noise, then losses would decrease the visibility. The robustness against losses in combination with its time-resolving capabilities with subpicoseconds accuracy, makes Hong-Ou-Mandel interferometry a useful technique.\\

With our setup we should be able to observe Hong-Ou-Mandel interference, as figure \ref{theory_HOM_fig_2} points out. It should be noted that for the sole purpose of performing a Hong-Ou-Mandel experiment, a type-I non-linear crystal is recommended above a type-II non-linear crystal for creating photon pairs. With a type-I non-linear crystal, the two created photons will be polarized along the same axis and hence experience the same dispersion of the non-linear crystal. This means that the spectrum for the trigger and idler photons will be identical and hence these photons are indistinguishable. In this case the visibility of Hong-Ou-Mandel interference will not depend on the pump bandwidth\cite{WPGrice_spectral_info_SPDC}. Hong-Ou-Mandel interferences with visibilities over 95\% using type-I non-linear crystals have been reported\cite{MTanida_good_HOM_SPDC}.  However, as mentioned in previous chapter, we are interested in performing many diverse experiments. For most of these experiments a type-II non-linear crystal is better than a type-I non-linear crystal. 

\section{Detection scheme}

\begin{figure}[tb!]
\centering
\includegraphics[scale=1.2]{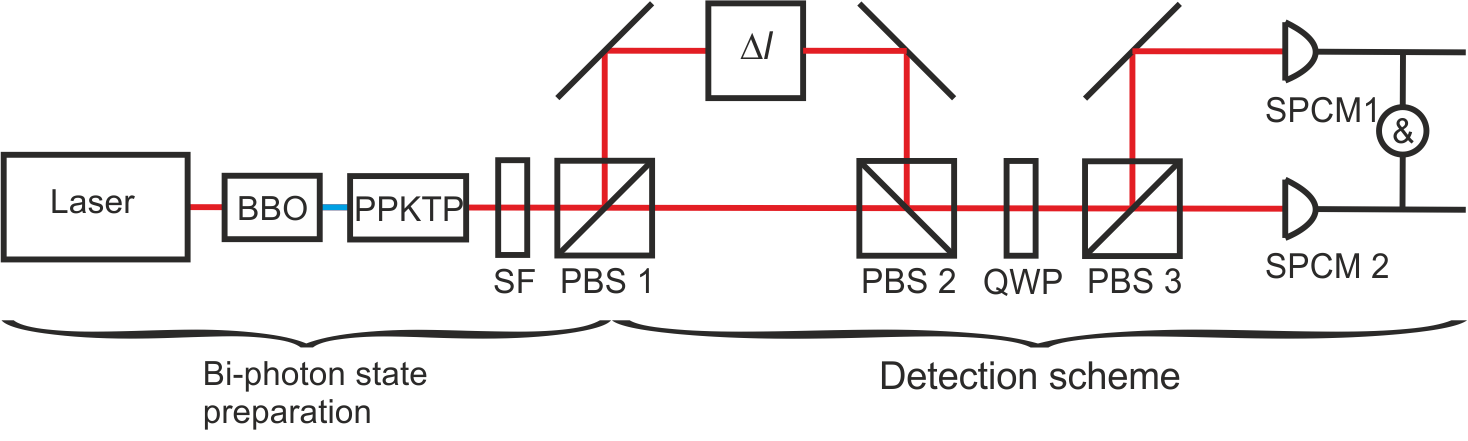}
\caption{\textbf{Simplified setup for the Hong-Ou-Mandel experiment.} A laser, BBO non-linear crystal (BBO) and a periodically poled KTP non-linear crystal (PPKTP) create entangled-photon pairs. The photons of these pairs are separated with a polarizing beam splitter (PBS 1). The reflected photons are sent to a variable delay ($\Delta l$) and afterwards recombined with the transmitted photons at PBS 2. A quarter-wave plate (QWP) is used to let the modes interfere at PBS 3. Coincidences are measured on two single-photon counting modules (SPCM 1 and SPCM 2).} \label{simplified_setup_HOM}
\end{figure}

An illustration of the setup used for the Hong-Ou-Mandel experiment is presented in figure \ref{simplified_setup_HOM}.
The left part of the setup (up to PBS 1) is used for the required bi-photon preparation, as discussed in the previous chapter. A spectral filter (SF) is added after the PPKTP non-linear crystal in order to make the photon spectra more identical. The photons reflected from polarizing beam splitter PBS 1 are send to a variable delay ($\Delta l$) and then recombined with the conjugate photons at a second polarizing beam splitter (PBS 2). With a quarter-wave plate (QWP) we can let the orthogonal polarized modes of the photons interfere with each other at a polarizing beam splitter (PBS 3). The output ports of PBS 3 are monitored with two single-photon counting modules (SPCM 1 and SPCM 2). The advantage of using a quarter-wave plate in combination with a polarizing beam splitter compared to a 50:50 beam splitter, is that we can also decide not to let both modes interfere with each other. This allows us to switch between a Hong-Ou-Mandel experiment and directly measuring photon rates for individual modes. A detailed detection scheme of the Hong-Ou-Mandel setup can be found in appendix \ref{HOM_setup_appendix}\\

\section{Measured Hong-Ou-Mandel interference}

\begin{figure}[!tb]
\centering
\subfloat[]{\includegraphics[scale=0.6]{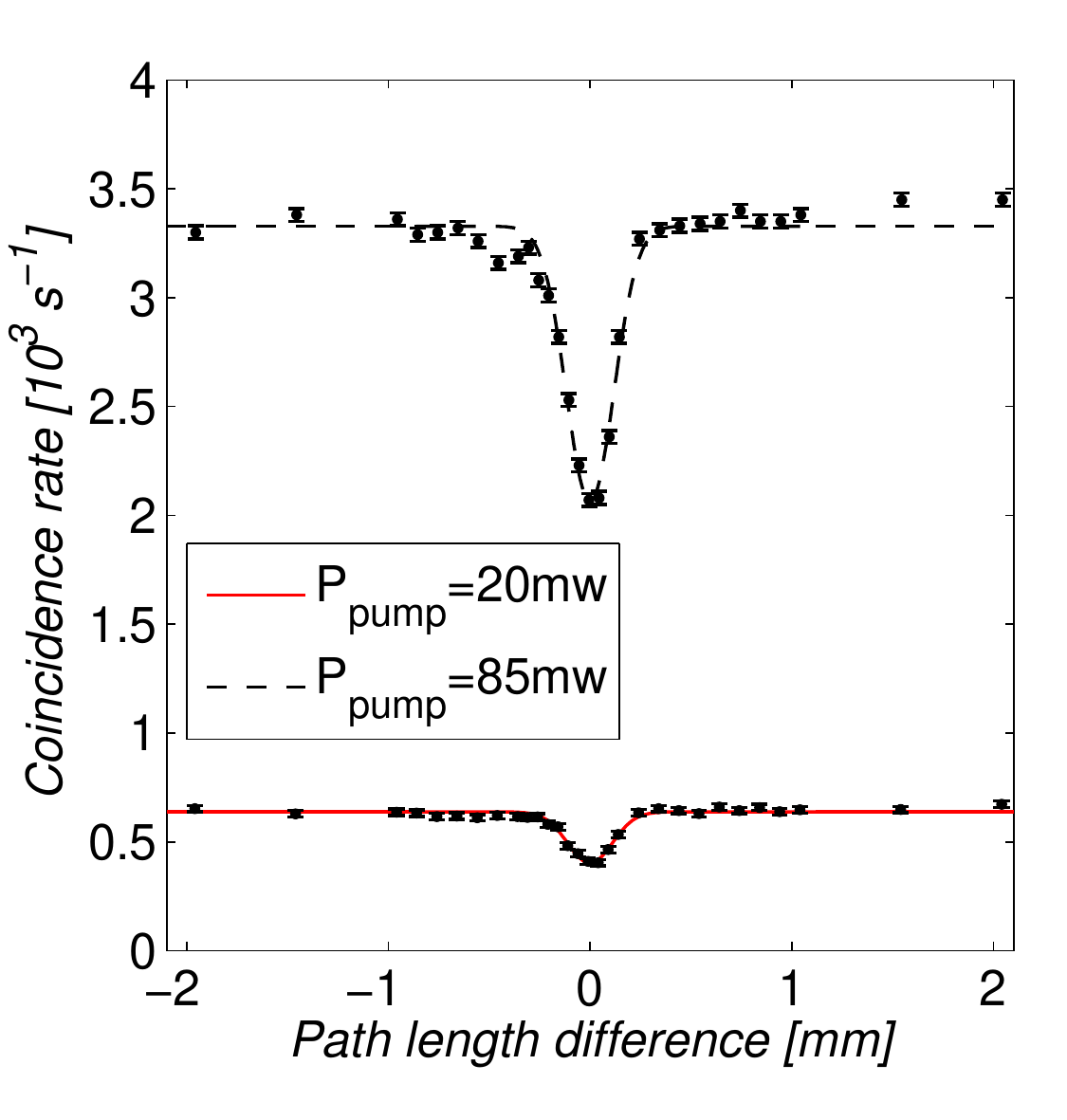}\label{data_HOM_fig_1}} 
\subfloat[]{\includegraphics[scale=0.6]{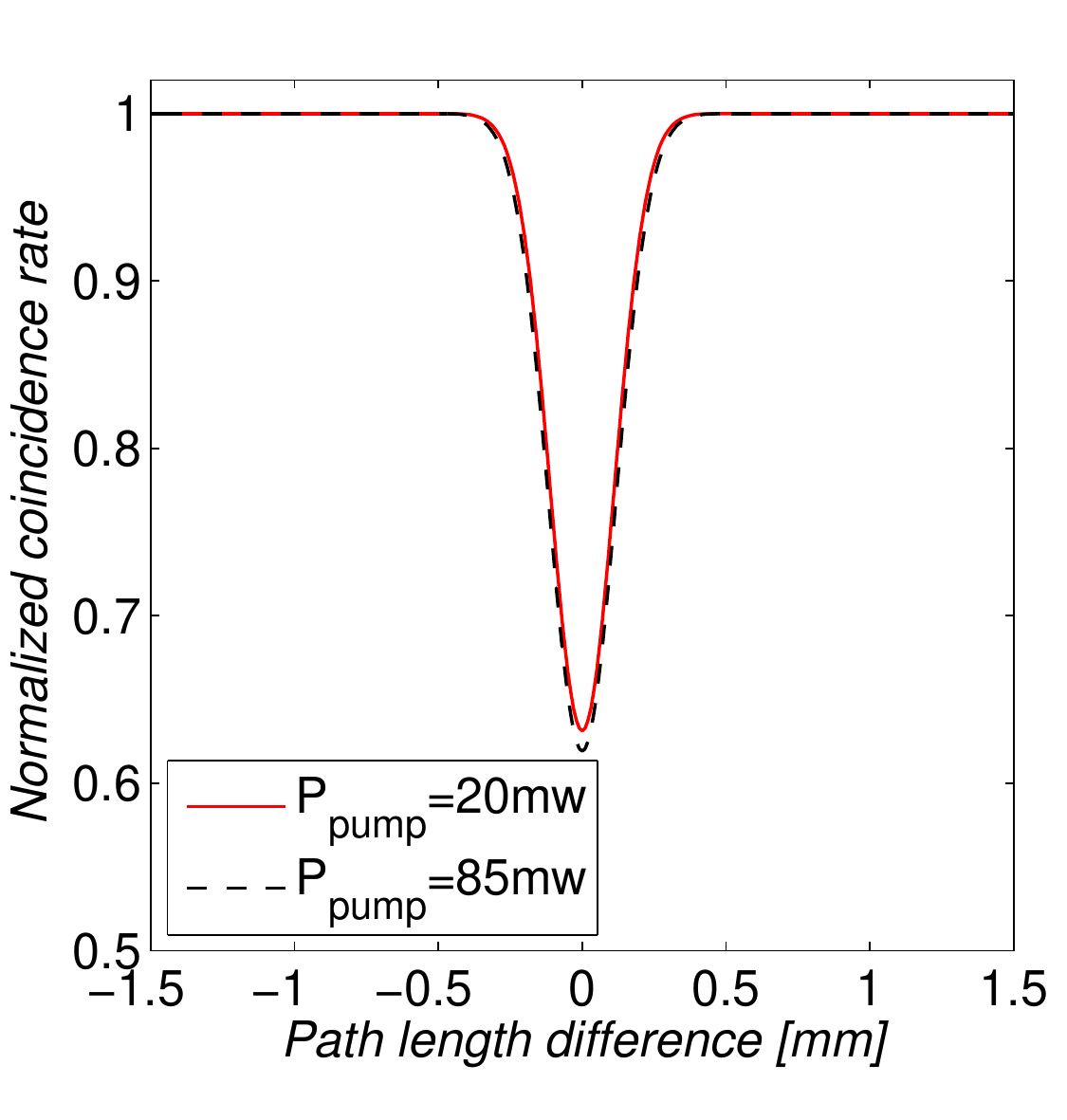}\label{data_HOM_fig_2}} 
\caption{\textbf{Experimental results of Hong-Ou-Mandel interference.} \textbf{(a)} The measured coincidence rate as function of the relative path length difference is presented for a pump power of 85 mW and 20 mW. The lines drawn in the figure are Gaussian fits. \textbf{(b)} The Gaussian fits of figure \textbf{(a)} are normalized as a way to compare them.}
\label{HOM_data}
\end{figure}

Our Hong-Ou-Mandel interference results are shown in figure \ref{HOM_data}.
In figure \ref{data_HOM_fig_1} we show our measured coincidence rate as function of the path length difference between the two arms ($\Delta l$) with errorbars indicating the maximum fluctuations observed during the measurement. We observe a pronounced dip when the path length difference is zero, demonstrating the presence of Hong-Ou-Mandel interference. We fitted our data points in figure \ref{data_HOM_fig_1} with a Gaussian function as a visual aid, which is commonly done\cite{CKHong_first_HOM,AMSteinberg_photon_glass_HOM,MTanida_good_HOM_SPDC}. The Gaussian fits of figure \ref{data_HOM_fig_1} are copied in figure \ref{data_HOM_fig_2} and normalized as a way to compare the Hong-Ou-Mandel interference at different pump powers. From the Gaussian fits in figure \ref{data_HOM_fig_2} it is easy to read that the visibility (depth of the dip) is about 38\% and has a full-width-at-half-maximum of approximately 0.23 mm. As mentioned earlier, the visibility indicates how indistinguishable the created photons are, while the width is an indication for the the spectrum of the photons. From figure \ref{theory_HOM_fig_2} we predicted the visibility to be about 22\% and the full-width-at-half-maximum to be around 1.28 mm. The increased visibility in the experiment compared to what we expected, means that the observed photons had spectra more identical to each other than predicted. As figure \ref{theory_HOM_fig_2} shows, spectrally filtering the two-photon spectrum after the PPKTP crystal increases the width of the Hong-Ou-Mandel interference. However, the measured width is narrower than what we expected. This means that the discrepancy between figures \ref{data_HOM_fig_2} and \ref{theory_HOM_fig_2} is mainly explained due to the used non-linear crystal properties in our model not being exactly equal to the properties of the crystal used in our experiment. Nevertheless, we are very satisfied to observe Hong-Ou-Mandel interference at count rates where short integration times ($<1$ s) are sufficient.\\ 

From figure \ref{data_HOM_fig_2} it also becomes clear that the Hong-Ou-Mandel interference does not depend significantly on the pump power. The visibility is 1\% higher for a pump power of 85 mW, which is comparable to the measured noise. The output of Hong-Ou-Mandel interference for the two-photon Fock state can be calculated like we did for the single-photon Fock state in equation (\ref{HOM_output_eq}). In this case the input state is given by $\mid \psi _{\rm{in}} \rangle = \mid 2_02_10_20_3 \rangle = \frac{1}{2} \hat{a}_0 ^\dag \hat{a}_0 ^\dag \hat{a}_1 ^\dag \hat{a}_1 ^\dag \mid 0_00_10_20_3 \rangle$. From this, we can derive
\begin{eqnarray}
\mid \psi _{\rm{out}} \rangle &=&  \frac{1}{2} \frac{1}{\sqrt{2}} \left( \hat{a}_2 ^\dag +i\hat{a}_3 ^\dag \right) \frac{1}{\sqrt{2}} \left( \hat{a}_2 ^\dag + i\hat{a}_3 ^\dag \right)  \frac{1}{\sqrt{2}} \left( \hat{a}_3 ^\dag + i\hat{a}_2 ^\dag \right)  \frac{1}{\sqrt{2}} \left( \hat{a}_3 ^\dag + i\hat{a}_2 ^\dag \right) \mid 0_00_10_20_3 \rangle \nonumber\\
&=& -\frac{1}{8}\left( \hat{a}_2 ^\dag \hat{a}_2 ^\dag \hat{a}_2 ^\dag  \hat{a}_2 ^\dag + 2 \hat{a}_2 ^\dag  \hat{a}_2 ^\dag  \hat{a}_3 ^\dag  \hat{a}_3 ^\dag + \hat{a}_3 ^\dag  \hat{a}_3 ^\dag  \hat{a}_3 ^\dag  \hat{a}_3 ^\dag  \right) \mid 0_00_10_20_3 \rangle \nonumber\\
&=& - \sqrt{\frac{3}{8}} \mid 0_00_14_20_3 \rangle - \frac{1}{2} \mid 0_00_12_22_3 \rangle - \sqrt{\frac{3}{8}} \mid 0_00_10_24_3 \rangle  \label{HOM_output_2_eq}
\end{eqnarray}
The result of this calculation is that the two-photon Fock state gives a quarter of the time coincidence counts. In chapter \ref{photon_source_chapter} we estimated that 1.5\% of the total observed counts are two-photon Fock states. If we take this into account for a Hong-Ou-Mandel interference visibility of 38\% when only the single-photon Fock state is present, then we would expect that the visibility would decrease approximately 0.3\% due to the presence of the two-photon Fock state. This means that the visibility of the Hong-Ou-Mandel interference is less for higher pump powers than for lower pump powers. We do not observe this in our data because of experimental noise. Thus we can conclude that the possibility of creating the two-photon Fock state in our setup has no significant influence on the observed Hong-Ou-Mandel interference.

\section{Conclusions of the Hong-Ou-Mandel experiment}

Hong-Ou-Mandel interference is a quantum interference effect where two photons impinge on a 50:50 beam splitter from different ports and leave the beam splitter always together. By placing two single-photon counting modules at the output ports of this beam splitter, we observed that the coincidence rate of the two counting modules can be decreased by 38\% as a result of quantum interference. Furthermore we observed that the possibility of creating the two-photon Fock state in our setup has no observable meaningful influence on the Hong-Ou-Mandel interference.\\

We expect that the observed coincidence rates can be lowered at least an order of magnitude before we will need to integrate the coincidence counts over more than 10 seconds, meaning that we can afford more losses. It is anticipated that we would be able to observe Hong-Ou-Mandel interference even with 99\% additional losses for one of the input modes of the 50:50 beam splitter. 

\appendix 

\chapter{Outline of the laser} \label{laser_appendix}

\begin{figure}[tb!]
\centering
\includegraphics[scale=0.5]{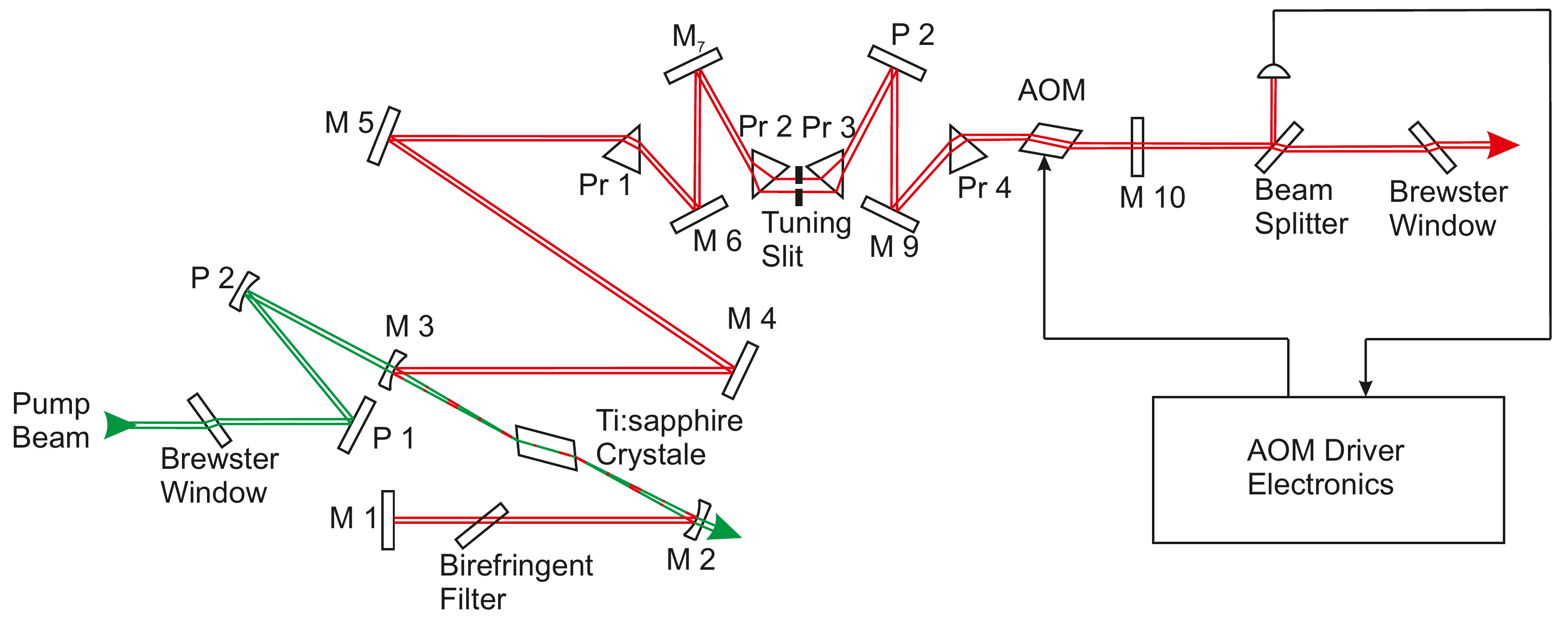}
\caption{\textbf{Overview of the Ti:Sapphire laser.} Pump light is focused inside a titanium sapphire crystal, creating the inversion necessary for lasing. Prisms Pr 1-4 compensate chirp acquired by pulses in the cavity. Picture adapted from reference \cite{Tsunami_manual}.} \label{Tsunami_laser}
\end{figure}

The titanium sapphire laser that we used is schematically given in figure \ref{Tsunami_laser}. 
Continuous-wave light at a wavelength of 532 nm enters the laser and is focused with mirrors P 1 and P 2 in the titanium sapphire crystal, which is the source of the lasing. Mirrors M 2 and M 3 can be used to overlap the cavity (lasing) mode with the pump profile inside the titanium sapphire crystal. Mirrors M 6 to M 9 in combination with prisms Pr 1 to Pr 4 are implemented to compensate (positive) chirp acquired by pulses in the cavity, making sure the output pulses are Fourier transform limited\cite{Tsunami_manual}\footnote{Fourier transform limited means that all wavelengths contributing to pulsed lasing are all maximally contributing to the peak intensity of the pulses. In this way a minimum temporal pulse width arises.}. Between prisms Pr 2 and Pr 3 the lasing wavelengths are spatially separated, allowing to choose the center wavelength and the spectral width using a slit. Mirrors M 1 and M 10 are the cavity mirrors that allow fine tuning of the cavity, making it possible to overlap the cavity mode optimally with the pump beam.\\
 
Lasing from the titanium sapphire crystal can occur in continuous wave or pulsed mode. If the alignment of the laser cavity is good, the power-depended part of the refractive index of the titanium sapphire crystal can result in pulsed lasing\cite{Tsunami_manual}. In this case, high peak intensities are amplified more by the crystal than low intensities, strengthening pulsed lasing. An acoustical optical modulator (AOM) (in combination with a photodiode monitoring the lasing) is used to aid this process.\\

We typically use 8.25 Watt light at a wavelength of 532 nm for creating the inversion necessary for lasing. This gives an average output power of the laser of approximately 750 mW. The center wavelength of the laser light is 790 nm and the full-width-at-half-maximum of the spectrum is typically around 3 nm. The corresponding temporal pulse width as measured with an autocorrelator is around 0.4 ps. 

\chapter{Detailed overview of the setup} \label{detailed_setup_appendix}

\section{Quantum light source setup} \label{single_photon_setup_appendix}

\begin{figure}[tb!]
\centering
\includegraphics[scale=1.2]{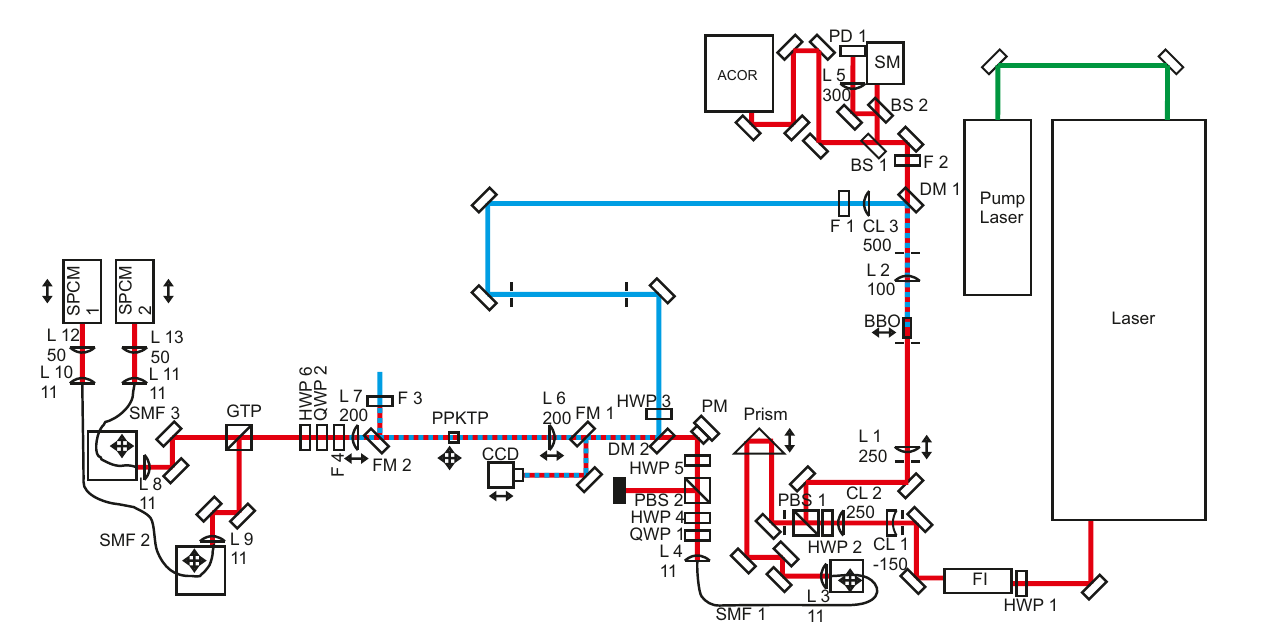}
\caption{\textbf{Detailed overview of the quantum light source.} Laser light is frequency doubled with a BBO non-linear crystal (BBO). The frequency-doubled light is used to pump a periodically poled KTP non-linear crystal (PPKTP), producing photon pairs. The photons of the created photon pairs are split with a Glan-Taylor polarizer (GTP) and detected separately at single-photon counting modules (SPCM 1 and SPCM 2).} \label{full_setup}
\end{figure}

A detailed schematic of the experimental setup can be found in figure \ref{full_setup}. 
In this section we describe the quantum light source as used in chapter \ref{photon_source_chapter}. Directly behind the laser a Faraday isolator (FI) is placed that allows light to pass through in only one direction. We found that without a Faraday isolator, reflections from different optical components in the setup could disrupt mode-locking of the laser. The laser beam is guided through two cylindrical lenses (CL 1 and CL 2) which compensate the ellipticity of our beam. With a polarizing beam splitter (PBS 1) we split our bundle into two parts and in combination with a half-wave plate (HWP 2) we can choose in what ratio the beam intensity is divided into these two paths. The two paths arising from the polarizing beam splitter are called the pump and seed path.\\

\subsection{Seed path} \label{seed_path_appendix}

The seed path is used as an alignment tool. The seed path contains a movable prism (Prism) which makes it possible to match the path length to the pump path length. The seed bundle is coupled into a single-mode fiber (SMF 1), providing us with a very good spatial mode quality. With polarizing optics (HWP 4-5, QWP 1 and PBS 2) directly behind the output of this single-mode fiber (L 4), we can obtain a well-defined linear polarization and tune the intensity of the fiber output without adjusting anything before the fiber. A mirror mounted on a piezoelectric element (PM) is used for alignment purposes and with a dichroic mirror (DM 2) the seed and pump bundle are superimposed on the PPKTP crystal (PPKTP).\\

\subsection{Pump path}

The purpose of the pump path is to create photon pairs at the PPKTP crystal. After leaving the polarizing beam splitter (PBS 1) the pump bundle is focused on a 5 mm long BBO crystal (BBO) using a 250 mm lens (L 1). Transmitted light from the BBO is collimated using a 100 mm lens (L 2) and spectrally filtered using a dichroic mirror (DM 1). Laser light is transmitted by this mirror and can be used to monitor the laser performance by means of an autocorrelator (ACOR), a spectrometer (SM) and a photodiode (PD 1). \\

Frequency-doubled light from the BBO is reflected by DM 1 and spectrally filtered (F 1: two Semrock FF01-440/SP-25) in order to remove remaining laser light. Also a cylindrical lens with a focal length of 500 mm  (CL 3) is used to compensate for astigmatism of the frequency-doubled light. The source of this astigmatism is expected to be the angle dependence of the frequency doubling. Typically after the filter and the cylindrical lens, we measure a power of 140 mW. The frequency-doubled light is passing through two irises clipping the beam as a way to spatially filter the beam. Eventually the pump beam is focused into the PPKTP crystal. With a half-wave plate (HWP 3) the polarization of the frequency-doubled light is matched to the orientation of the PPKTP.\\

\subsection{After the PPKTP crystal}

Both the seed and pump path pass through the PPKTP crystal and are collimated with a lens (L 7), after which the frequency-doubled light is removed with a spectral filter (F 4: Semrock BLP01-635R-25). A half-wave plate (HWP 6) and a quarter-wave plate (QWP 2) allow for compensating birefringence of the PPKTP crystal and for optionally mixing orthogonal polarization states at the Glan-Taylor polarizer (GTP). The pump beam creates photon pairs at the PPKTP, which are separated at the GTP. The created photons are coupled into single-mode fibers (SMF 2-3) of which the outputs are focused on two separate single-photon counting modules (SPCM 1 and SPCM 2: PerkinElmer SPCM-AQRH-14 and PerkinElmer SPCM-AQRH-13 respectively). The outputs of these single-mode fibers together with the SPCM's and all optics in between are placed into a separate box in order to reduce the number of counts due to stray light. 

\section{Hong-Ou-Mandel detection scheme} \label{HOM_setup_appendix}

\begin{figure}[tb!]
\centering
\includegraphics[scale=1.6]{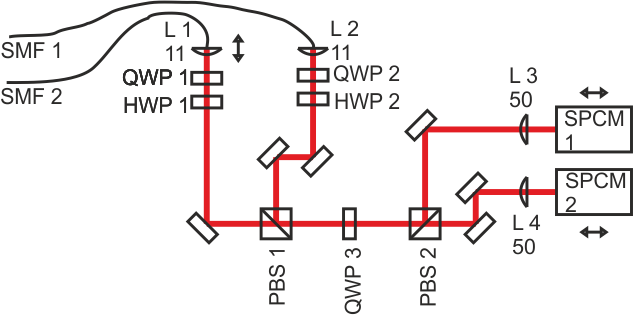}
\caption{\textbf{Detailed detection scheme of the setup used for the Hong-Ou-Mandel experiment.} Two photons are superimposed on a polarizing beam splitter (PBS 1). A quarter-wave plate (QWP 3) in combination with a polarizing beam splitter (PBS 2) act as a 50:50 beam splitter. The path length difference between the two photons can be controlled by moving output lens L 1.} \label{detailed_setup_HOM}
\end{figure}

In figure \ref{detailed_setup_HOM} we present a detailed overview of the detection scheme used for the Hong-Ou-Mandel experiment (see chapter \ref{HOM_chapter}). 
Two separate photons enter the detection scheme through single mode fibers (SMF 1 and SMF 2) of which the output lenses are illustrated in figure \ref{detailed_setup_HOM} by L 1 and L 2. Output lens L 1 is placed on a translation stage making it possible to control the path length difference with respect to the other photon. Quarter-wave and half-wave plates (QWP 1, QWP 2, HWP 1 and HWP 2) are included to control the polarization. Both photons are superimposed on a polarizing beam splitter (PBS 1). The photons after PBS 1 have orthogonal polarizations. The combination of QWP 3 and PBS 2 act as a 50:50 beam splitter for the photons. The output arms of this 50:50 beam splitter are monitored with two single-photon counting modules (SPCM 1 and SPCM 2). These counting modules are placed on translation stages (indicated with arrows in figure \ref{detailed_setup_HOM}) so that they can be placed in the foci of lenses L 3 and L 4. 

\subsection{Spectral filter for the Hong-Ou-Mandel experiment} \label{spectral_filter_section}

For the purpose of making a Hong-Ou-Mandel interferometer, we desired a spectral filter with a bandwidth below 2 nm at a center wavelength of 790 nm. We have decided to fabricate a Fabry-Perot etalon and use it as a spectral filter. This etalon has the ability to tune the center wavelength and has a high transmission for the center wavelength. The bandwidth of our laser pulses demands that the free spectral range, the spectral distance between two consecutive transmission peaks of such an etalon should be larger than 5 nm. Theoretically, the free spectral range (in air) is given by\cite{fundamentals_of_photonics}
\begin{figure}[tb!]
\centering
\subfloat[]{\includegraphics[scale=1.2]{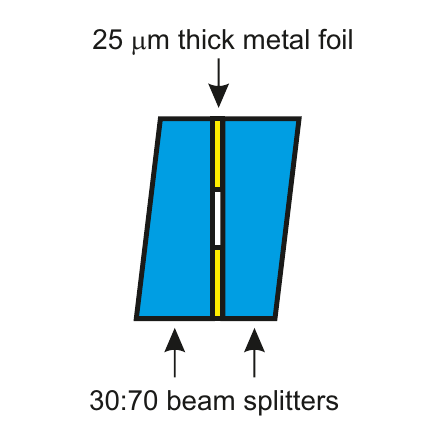}\label{spectral_fitler_fig_1}} 
\subfloat[]{\includegraphics[scale=0.6]{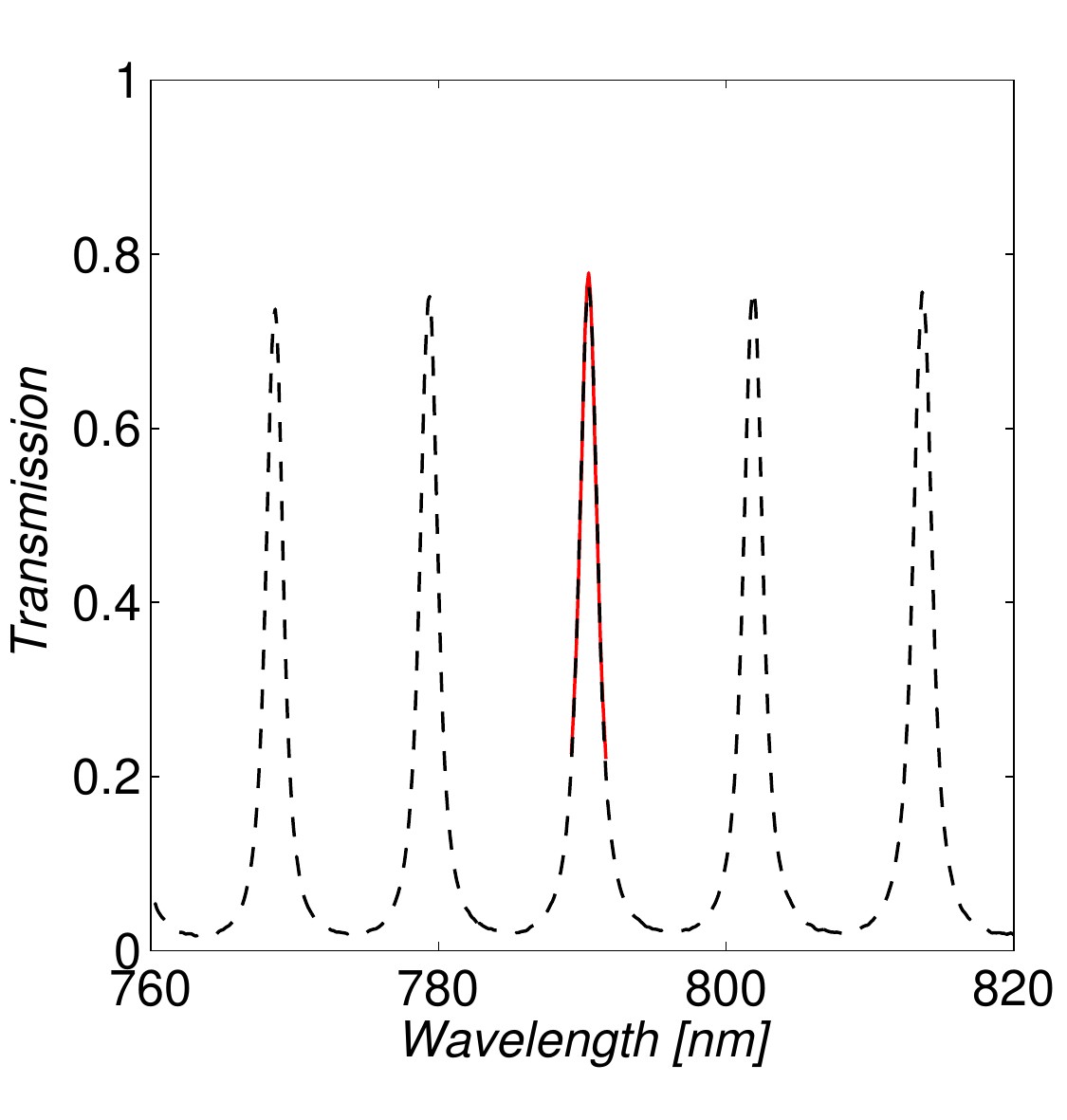}\label{spectral_fitler_fig_2}} 
\caption{\textbf{Spectral filter for Hong-Ou-Mandel experiment.} \textbf{(a)} Overview of how the spectral filter was fabricated. \textbf{(b)} Measured spectrum of our fabricated filter. The red line is a Lorentzian fit.} \label{spectral_fitler_figs}
\end{figure}
\begin{equation}
\Delta \lambda _{F} = \frac{\lambda ^2}{2d}
\end{equation}
where $d$ is the spacing between the mirrors of the etalon. The spectral width is defined as
\begin{equation}
\delta \lambda = \frac{\Delta \lambda _{F}}{\mathcal{F}}
\end{equation}
where $\mathcal{F}$ is the Finesse of the etalon given by
\begin{equation}
\mathcal{F}=\frac{\pi \sqrt{R}}{1-R}
\end{equation}
in which $R$ represents the average intensity reflectivity of the mirrors of the etalon. For our etalon we used two 30:70 (transmission:reflectivity) beam splitters with metal foil of 25 $\mu$m thick in between as spacer, see figure \ref{spectral_fitler_fig_1}. 
Theoretically this gives us a free spectral range of 12.4 nm and a spectral width of 1.4 nm. Figure \ref{spectral_fitler_fig_2} shows the spectrum of our filter when a transmission peak is set to 790 nm. Fitting the peak transmission at 790 nm with a Lorentzian function gives a spectral width of 1.5 nm and a peak transmission of 0.76. Thereby we observe a spectral range of 11.4 nm in figure \ref{spectral_fitler_fig_2}. We measured the spectrum using a broadband white light source (Fianium SC-450) and an interferometer (BioRad FTS-6000). For the Hong-Ou-Mandel interference measurements we implemented this spectral filter directly after the PPKTP crystal so that all the created photons are spectrally filtered.


\begin{thebibliography}{99}
\bibitem{DCBurnham_parametric_photon_pairs}
D. C. Burnham and D. L. Weinberg,
``Observation of Simultaneity in Parametric Production of Optical Photon Pairs'',
Physical Review Letters \textbf{25}, 84-87 (1970).

\bibitem{SACastelletto_heralded_photon_source}
S. A. Castelletto and R. E. Scholten,
``Heralded single photon sources: a route towards quantum communication technology and photon standards'',
The European Physical Journal Applied Physics, \textbf{41}, 181-194 (2008).

\bibitem{PGKwiat_new_engtangled_source}
P. G. Kwiat, K. Mattle, H. Weinfurter and A. Zeilinger,
``New High-Intensity Source of Polarization-Entangled Photon Pairs'',
Physical Review Letters, \textbf{75}, 4337-4341 (1995).

\bibitem{NJain_bridge_single_squeezed}
N. Jain, S. R. Huisman, E. Bimbard and A. I. Lvovsky,
``A bridge between the single-photon and squeezed-vacuum states'',
Optics Express, \textbf{18}, 18254-18259 (2010).

\bibitem{AZavatta_single_cohterent_state}
A. Zavatta, S. Viciani and M. Bellini,
``Single-Photon-Added Coherent States of Light'',
Science, \textbf{306}, 660-662 (2004).

\bibitem{EBimbard_state_engineering}
E. Bimbard, N. Jain, A. MacRae and A. I. Lvovsky,
``Quantum-optical state engineering up to the two-photon level'',
Nature Photonics, \textbf{4}, 243-247 (2010).

\bibitem{fundamentals_of_photonics}
B. E. A. Saleh and M. C. Teich,
``Fundamentals of Photonics'',
Wiley, edition 2 (2007).

\bibitem{WPGrice_spectral_info_SPDC}
W. P. Grice and I. A. Walmsley,
``Spectral information and distinguishability in type-II down-conversion with a broadband pump'',
Physical Review A, \textbf{56}, 1627-1634 (1997).

\bibitem{GSBuller_single_photon_generation}
G. S. Buller and R. J. Collins,
``Single-photon generation and detection'',
Measurment Science and Technology, \textbf{21}, 012002 (2010).

\bibitem{MKeller_single_photons_ion_trap}
M. Keller, B. Lange, K. Hayasaka, W. Lange and H. Walther,
``Continuous generation of single photons with controlled waveform in an ion-trap cavity system'',
Nature, \textbf{431}, 1075-1078 (2004).

\bibitem{CKurtsiefer_diamond_photons}
C. Kurtsiefer, S. Mayer, P. Zarda and H. Weinfurter,
``Stable Solid-State Source of Single Photons'',
Physical Review Letters, \textbf{85}, 290-293 (2000).

\bibitem{PMichler_single_quantum_dot}
P. Michler, A. Imamoglu, M. D. Mason, P. J. Carson, G. F. Strouse and S. K. Buratto,
``Quantum correlation among photons from a single quantum dot at room temperature'',
Nature, \textbf{406}, 968-970 (2000).

\bibitem{SStrauf_many_single_photons}
S. Strauf, N. G. Stoltz, M. T. Takher, L. A. Coldren, P. M. Petroff and D. Bouwmeester,
``High-frequency single-photon source with polarization control'',
Nature Photonics, \textbf{1}, 704-708 (2007).

\bibitem{HHansen_PhD_thesis}
H. Hansen,
``Generation and Characterization of New quantum States of the Light Field'',
Doctoral thesis (2000).

\bibitem{NJain_MSc_thesis}
N. Jain,
``Quantum optical state engineering at the few-photon level'',
Master thesis (2009).

\bibitem{Simon_internship_report}
S. R. Huisman,
``Instant single-photon-Fock state tomography'',
Internship Report (2009).

\bibitem{APD_sheet}
PerkinElmer,
``SPCM-AQRH Single Photon Couting Module'',
Datasheet (2007).

\bibitem{OSvelto_principles_lasers}
O. Svelto,
``Principles of Lasers'',
Plenum Press, New York \& London, edition 4 (1998).

\bibitem{Tsunami_manual}
Spectra-Physics,
``Tsunami Mode-locked Ti:sapphire Laser'',
User's manual (2002).

\bibitem{SRHuisman_single_photon_tomography}
S. R. Huisman, N. Jain, S. A. Babichev, F. Vewinger, A. N. Zhang, S. H. Youn and A. I. Lvovsky,
``Instant single-photon Fock state tomography'',
Optics Letters, \textbf{34}, 2739-2741 (2009).

\bibitem{CKHong_first_HOM}
C. K. Hong, Z. Y. Ou and L. Mandel,
``Measurement of Subpicosecond Time Intervals between Two Photons by Interference'',
Physical Review Letters, \textbf{59}, 2044-2046 (1987).

\bibitem{quantum_optics_Fox}
M. Fox,
``Quantum Optics An Introduction'',
Oxford University Press (2007).

\bibitem{JCGarrison_quantum_optics}
J. C. Garrison and R. Y. Chiao,
``Quantum Optics'',
Oxford University press (2008).

\bibitem{AMSteinberg_photon_glass_HOM}
A. M. Steinberg, P. G. Kwiat and R. Y. Chiao,
``Dispersion Cancellation in a Measurement of the Single-Photon Propagation Velocity in Glass'',
Physical Review Letters, \textbf{68}, 2421-2424 (1992).

\bibitem{AMSteinberg_photon_tunneling_HOM}
A. M. Steinberg, P. G. Kwiat and R. Y. Chiao,
``Measurement of the Single-Photon Tunneling Time'',
Physical Review Letters, \textbf{71}, 708-711 (1993).

\bibitem{DJPapoular_optical_tunneling_HOM}
D. J. Papoular, P. Clad\'e, S. V. Polyakov, C. F. McCormick, A. L. Migdall and P. D. Lett,
``Measuring optical tunneling times using a Hong-Ou-Mandel interferometer'',
Optics Express, \textbf{16}, 16005-16012 (2008).

\bibitem{BZhao_robust_creation}
B. Zhao, Z.-B. Chen, Y.-A. Chen, J. Schmiedmayer and J.-W. Pan,
``Robust Creation of Entanglement between Remote Memory Qubits'',
Physical Review Letters, \textbf{98}, 240502 (2007).

\bibitem{quantum_optics_Gerry_Knight}
C. C. Gerry and P. L. Knight,
``Introductory Quantum Optics'',
Cambridge University press (2006).

\bibitem{ANBoto_quantum_lithography}
A. N. Boto, P. Kok, D. S. Abrams, S. L. Braunstein, C. P. Williams and J. P. Dowling,
``Quantum Interferometric Optical Lithography: Exploiting Entanglement to Beat the Diffraction Limit'',
Physical Review Letters, \textbf{85}, 2733-2736 (2000).

\bibitem{YKawabe_quantum_diffraction_limit}
Y. Kawabe, H. Fujiware, R. Okamoto, K. Sasaki and S. Takeuchi,
``Quantum interference fringes beating the diffraction limit'',
Optics Express, \textbf{15}, 14244-14250 (2007).

\bibitem{TNagata_beating_quantum_limit}
T. Nagata, R. Okamoto, J. L. O'Brien, K. Sasaki and S. Takeuchi,
``Beating the Standard Quantum Limit with Four-Entangled Photons'',
Science, \textbf{316}, 726-729 (2007).

\bibitem{IAfek_high_NOON}
I. Afek, O. Ambar and Y. Silberberg,
``High-N00N States by Mixing Quantum and Classical Light'',
Science, \textbf{328}, 879-881 (2010).

\bibitem{GGhosh_dispersion_KTP}
G. Ghosh,
``Temperature Dispersion in KTP for Nonlinear Devices'',
IEEE Photonics Technology Letters, \textbf{7}, 68-70 (1995).

\bibitem{MTanida_good_HOM_SPDC}
M. Tanida, R. Okamoto and S. Takeuchi,
``Highly indistinguishable heralded single-photon sources using parametric down conversion'',
Optics Express, \textbf{20}, 15275-12585 (2012).



\end{thebibliography}
\end{document}